\documentclass[twocolumn, aps, prl,showkeys, superscriptaddress, showpacs, amsmath,amssymb]{revtex4}
\usepackage[T1]{fontenc} 
\usepackage{graphicx}
\usepackage{dcolumn}
\usepackage{bm}
\usepackage{color}

\begin{document}


\title{Statistical properties of one dimensional attractive Bose gas}

\author{Przemys\l aw Bienias}
\affiliation{
College of Inter-Faculty Individual
Studies in Mathematics and Natural
Sciences, University of Warsaw, ul. \.Zwirki i Wigury 93, 02-089
Warszawa, Poland
}
\affiliation{
Faculty of Physics, Warsaw University, ulica Ho\.{z}a 69, PL-00-681, Warsaw, Poland
}

\author{Krzysztof Paw\l owski}
\affiliation{
Center for Theoretical Physics, Polish Academy of Sciences, Aleja Lotnik\'ow 32/46, 02-668 Warsaw, Poland
}
\affiliation{
5. Physikalisches Institut, Universit\"at Stuttgart, Pfaffenwaldring 57, 70550 Stuttgart, Germany
}

\author{Mariusz Gajda}
\affiliation{
Institute of Physics, Polish Academy of Sciences, Aleja Lotnik\'ow 32/46, 02-668 Warsaw, Poland
}

\author{Kazimierz Rz\k{a}\.{z}ewski}
\affiliation{
Center for Theoretical Physics, Polish Academy of Sciences, Aleja Lotnik\'ow 32/46, 02-668 Warsaw, Poland
}
\affiliation{Faculty of Mathematics and Sciences, Cardinal Stefan Wyszy\'nski University, ulica Dewajtis 5, 01-815, Warsaw, Poland}%
\affiliation{
5. Physikalisches Institut, Universit\"at Stuttgart, Pfaffenwaldring 57, 70550 Stuttgart, Germany
}

\date{\today}

\begin{abstract}
Using classical field approximation we present the first study of
statistical properties of one dimensional Bose gas with attractive
interaction. The canonical probability distribution is generated with the
help of a Monte Carlo method. This way we obtain not only the depletion of
the condensate with growing temperature but also its fluctuations. The
most important is our discovery of a reduced coherence length, the
phenomenon observed earlier only for the repulsive gas, known as
quasicondensation.
\end{abstract}

                             
                 
\maketitle

\section{Introduction}

Statistical properties of finite number of bosons confined by a trapping potential are
intensively studied ever since the very first experiments on dilute gas Bose-Einstein
condensates. Initially a full understanding has been reached of the statistics of the
confined ideal Bose gas \cite{grossman1996, politzer1996, gajda1997, navez1997, grossman1997, weiss1997}. 
In this context criticism of the most widely used grand
canonical ensemble has been raised. As it turns out it predicts unphysical fluctuations of
the number of condensed atoms. Interactions, an essential aspect of Bose atoms physics
make the problem of statistics rather complicated. In the simplest, although still academic
case, that of the box with periodic boundary condition, at least we know a priori what is a
condensate. It is the zero momentum component of the atomic field. In the box with
periodic boundary conditions we know more: we know analytically the spectrum and the
expressions for the collective Bogoliubov excitations. So, one can look at the statistics of
the gas as that of noninteracting bosonic quasiparticles. This determines the statistical
properties of the gas at least at low temperatures \cite{giorgini1996}. Extension of the method to higher
temperatures is already significantly more complicated. The Bogoliubov-Popov spectrum
depends on the actual number of the condensed atoms rather than on their total number.
Thus, itself is a fluctuating variable. Several papers attempted to take this into account \cite{svidzinsky2006, idziaszek2003}. 
Another serious problem is the assumption that quasiparticles do not
interact. In fact they do and they have a finite (temperature dependent) lifetime. In a
realistic harmonic trap we have no analytic formulae for the quasiparticles and also the
condensate degree of freedom is known only a posteriori since it is a function of an
interaction strength, a frequency of the harmonic trap, a number of atoms and a
temperature of the sample.
In two recent papers \cite{witkowska2009, bienias2011} we have shown how to overcome most of these
problems. To this end we proposed to use the classical field description of the system
generating the proper thermal equilibrium distribution using a classical Metropolis
algorithm \cite{metropolis1953}. 
Details of classical field approaches are presented in \cite{brewczyk2007, proukakis2008}.
It is the purpose of this Letter to report the first study of statistical properties of an
attractive Bose gas. Confined to a box it is unstable. Hence, this simple model makes no
sense for the attractive gas. The situation is different in a harmonic trap. Limited attractive
condensates do exist in two and three dimensions. The lithium-7 condensate was among
the first ones to be observed \cite{bradley1995} . It then has lead to creation of bright solitons
\cite{strecker2002, khaykovich2002}. In strictly 1D systems the effective repulsion of the kinetic energy makes
it stable for any number of particles. We therefore concentrate our attention on the one
dimensional, finite, attractive Bose gas trapped in a harmonic potential.
Thus the Hamiltonian of the system has the form:
\begin{eqnarray}
\nonumber H&=& \int \hat{\Psi}^{\dagger }(x)\left( \frac{p^2}{2m}+\frac{1}{2}m\,\omega_0^2 x^2  \right) \hat{\Psi}(x) dx +\\
& & +\,\frac{g}{2} \int \hat{\Psi}^{\dagger }(x) \hat{\Psi}^{\dagger }(x) \hat{\Psi}(x) \hat{\Psi}(x) ,
\label{eqn:hamiltonian}
\end{eqnarray}
where $\hat{\Psi}$ is the atomic field operator, $m$ is the mass of an atom, $\omega_0$ is the frequency of the
harmonic trapping potential and we are particularly interested in $g<0$. The case of positive
coupling constant was studied in detail in \cite{bienias2011}.

The classical fields approximation consists in replacement of the quantized atomic field by
a c-number wave function. This wave function is then conveniently expanded as a sum
over the harmonic oscillator wave functions and expansion coefficients $\alpha_n$ are classical
stochastic complex variables.
\begin{equation}
\Psi(x) = \sum _{n=0}^{n_{max}} \alpha_n \varphi_n(x) 
\label{eqn:clasField}
\end{equation}
where the oscillator eigenfunctions $\varphi_n$ are chosen to correspond to a harmonic oscillator of
the frequency $\omega$. Note that $\omega$ is not necessarily equal to $\omega_0$.

Thus the energy functional:
\begin{equation}
E\left[\Psi\right] = \int \Psi^{*}(x)\left( \frac{p^2}{2m}+\frac{1}{2}m\,\omega_0^2 x^2  \right) \Psi(x)  +\frac{g}{2}\int |\Psi(x)|^4 ,
\label{eqn:energyPsi}
\end{equation}
takes a form:
\begin{eqnarray}
\nonumber E\left( \left\{ \alpha_n \right\} \right) &=& \sum_{n=0}^{n_{max}} \hbar \omega n |\alpha_n |^2 + E_{int}\left( \left\{ \alpha_i \right\} \right) +\\
& &\sum_{n, n^{\prime}=0}^{n_{max}}\frac{1}{2} m \left(\omega_0^2-\omega^2 \right) \langle n|x^2|n^{\prime}\rangle \alpha^*_n\alpha_{n^{\prime}}
\label{eqn:energy}
\end{eqnarray}
with the interaction energy $E_{int}$ being a quartic form in the variables $\alpha_n$ and $\langle n | x^2| n^{\prime}\rangle$ is
the matrix element in the oscillator basis.

We are going to look at the statistics of a sample composed of $N$ atoms, so the amplitudes are subject to a
constraint:
\begin{equation}
\sum^{n_{max}}_{n=0}|\alpha_n|^2=N
\end{equation}
Having converted a quantum statistical physics problem into classical one we then define
a canonical equilibrium distribution of the amplitudes $\alpha_n$ as
\begin{equation}
P\left( \left\{ \alpha_i \right\} \right) \propto \exp \left[-\frac{E\left( \left\{ \alpha_i \right\} \right)}{k_B T}\right]
\label{eqn:probability}
\end{equation}
An efficient algorithm generating this probability distribution has been invented by
Metropolis \cite{metropolis1953}.

It is easy to check with the help of Gross-Pitaevskii equation imaginary time propagation
that the zero temperature attractive 1D Bose gas has a wave function which is very close
to a Gaussian. Its width shrinks with growing number of condensed atoms and with
growing absolute value of the negative coupling constant $g$. This is our guide to the optimal
choice of the frequency $\omega$ defining the actual oscillator base in the expansion of the atomic
classical field \eqref{eqn:clasField}. Thus this frequency becomes a function of the coupling and of the
temperature as they both determine the width of the condensate wave function. More
precisely: as we have shown in \cite{witkowska2009} the classical fields reproduce
faithfully the available analytically exact probability distribution of a harmonically confined
ideal Bose gas for the cut-off parameter $n_{max}$ satisfying:
\begin{equation}
n_{max} \hbar\omega_0 = k_BT
\end{equation}
where $T$ is the temperature of the gas. A natural generalization of this condition for a
weakly attractive 1D Bose gas is 
\begin{equation}
n_{max} \hbar\omega (N, T) = k_BT
\label{eqn:cutoff}
\end{equation}
where the frequency $ \omega (N, T)$ is chosen such that the corresponding Gaussian function
describes the ground state of the Gross-Pitaevskii equation for $N_0$ interacting atoms.
Strictly speaking $N_0$ should be chosen self consistently. In this Letter, however, we are
satisfied with $N_0$ being a number of condensed \textbf{ideal} gas atoms at given temperature. The
general rule for determination of the condensate wave function follows from the Onsager-
Penrose \cite{penrose1956} definition calling for the diagonalization of the one-particle density matrix
\begin{equation}
\rho_{i,j}=\langle\alpha_i^* \alpha_j\rangle=\sum_n\lambda_n\, \beta_i^*(n) \beta_j(n)
\label{eqn:rho}
\end{equation}
and identification of the condensate wave function with the eigenvector corresponding to
the leading eigenvalue. As a cross-check we verified that our identification of base indeed
yields the condensate wave function very close to the n=0 state.

\begin{figure}
\includegraphics[width= 8.6cm, angle=0, clip=true]{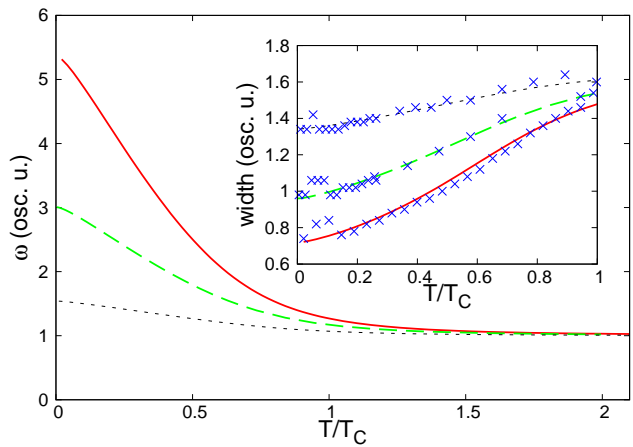}
\caption{(Color online)Effective frequency $\omega (N, T)$ for various values of the parameter $g$. The total number of atoms $N=500$.
Solid red line corresponds to $g=-0.01$, dashed green to $g=-0.0067$ and dotted black to $g=-0.003$.
In inset we compare the FWHM of the Gaussian with effective $\omega$ with the width of the condensate (crosses).
}
\label{fig:omega}
\end{figure}
It is clear that our choice of the frequency $\omega (N, T)$ makes it monotonically decrease with
the temperature and tend to that of the empty trap: $\omega_0$. In Fig. \ref{fig:omega} we illustrate this for several
values of the coupling $g$. In the inset we also confirm a consistency of our choice of the
base by comparing the width of the ground state of the Gross-Pitaevskii equation with the
width of the actual condensate wave function computed via diagonalization of the one-
particle density matrix. Small steps visible in the inset result from taking the solution of \eqref{eqn:cutoff}
for the cut-off as the nearest integer.

Multiparticle Hamiltonian with interactions dependent on the mutual distances between
atoms with the harmonic potential trapping has the center of mass of the system obey
interaction free Schr\"odinger equation decoupled from all other degrees of freedom.
Thus
this variable is entirely missing from any mean field approximation in which a nonlinear
Gross-Pitaevskii equation describes the system. This does not have important
consequences for the repulsive gas. In this case, the condensate is broader than the
ground state of the harmonic potential, so the uncertainty over the position of the center is
just a correction. The situation for attractive gas is different. Now, the condensate is
spatially squeezed and the shot-to-shot variation of the position of the center of mass is at
least as broad as the ground state of the empty harmonic potential \cite{gajda2001, gajda2006}. 
We therefore stress that our approach disregards this uncertainty, so the statistical
properties derived here pertain to the reference frame of the center of mass of the system.
Measurements in the center of mass system can meet some difficulties due to uncertainty
of position of the center of mass. However, a single shot simultaneous detection of many
particles resolves this problem \cite{gajda2006}.
\begin{figure}
\includegraphics[width= 8.6cm, angle=0, clip=true]{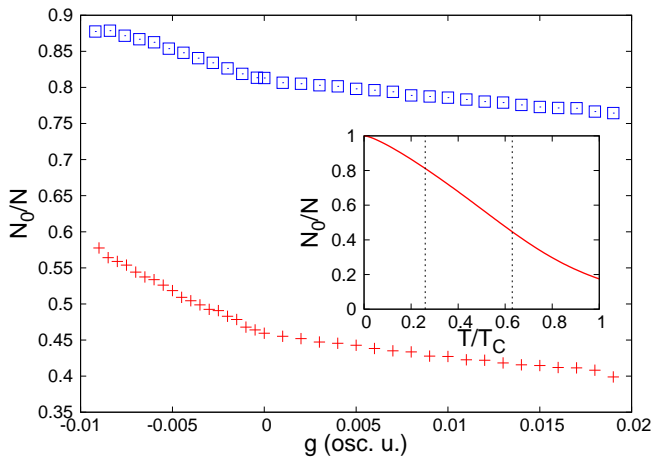}
\caption{(Color online) Number of atoms in the condensate as a function of the interaction strength $g$ for two fixed temperatures $T=0.26T_C$ (squares) and $T=0.63T_C$ (crosses).
These temperatures are indicated in the inset, where the decay of condensed atoms in the ideal gas is presented.
}
\label{fig:n0}
\end{figure}

In our classical field approximation effects of quantum fluctuations and quantum depletion
are missing. The method is suitable for weakly interacting Bose gas in the quantum
degenerate regime. Throughout this paper we use the oscillator units of position,energy
and temperature, $\sqrt{\frac{\hbar}{m \omega_0}} $, $\hbar \omega_0$ and $\frac{\hbar \omega_0}{k_B}$
respectively. Hence a dimensionless coupling $g$ is in units of $\sqrt{\frac{\hbar^3\omega_0}{m}}$.

All presented results are for $N=500$ atoms and the temperature is in units of the characteristic temperature indicating transition
to the quantum degenerate regime of a finite sample of an ideal Bose gas $N = T_C ln(2T_C)$
where $N$ is the total number of atoms \cite{ketterle1996}. 
In Fig. \ref{fig:n0} we plotted a number of condensed atoms as a function of the coupling constant $g$
for negative and also positive values (using the method explained in detail in \cite{bienias2011}) for
two temperatures, one very low ($T/T_C=0.26$) and the other $T/T_C=0.63$. Since the condensation depends on the local density of
particles, the effects of positive and negative interactions are opposite. The first one
swells the condensate reducing the local density with respect to the ideal gas while the
other shrinks it enhancing the condensation process. Thus the number of condensed
atoms is a monotonically decreasing function of the interaction parameter $g$ as we go from
the negative to the positive values.
\begin{figure}
\includegraphics[width= 8.6cm, angle=0, clip=true]{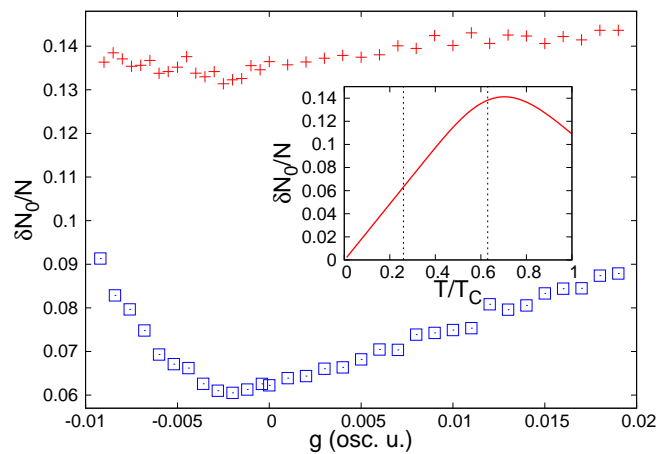}
\caption{(Color online) Dispersion of number of condensed atoms as a function of the interaction strength $g$ for two fixed temperatures $T=0.26T_C$ (squares) and $T=0.63T_C$ (crosses).
These temperatures are indicated in the inset, where the fluctuation of condensed atoms in the ideal gas is presented.
}
\label{fig:deltan0}
\end{figure}
In Fig. \ref{fig:deltan0} we present the relative fluctuations also as a function of changing sign $g$ and for
the same temperatures as in Fig. \ref{fig:n0}. General observation is that fluctuations of the
condensed atoms number grow with the strength of the interaction in both negative and
positive directions \footnote{For $T>0.75$ the fluctuations start to decrease, see Fig. 3 in \cite{bienias2011}}. An unexpected feature, however is that the minimal fluctuations are
observed not for $g=0$ but for tiny attractive interaction of $-0.0025$.

Perhaps the most interesting aspect of a 1D repulsive Bose gas predicted \cite{petrov2000}
and then observed experimentally \cite{dettmer2001, esteve2006} is the phenomenon of a
quasicondensation. It is a reduction of the coherence length to values smaller than the
length of the condensate above a certain characteristic temperature. In the recent Letter \cite{witkowska2011} the correlation length has
been related to the gray solitons formed during a rapid cooling process \cite{zurek2009}. We have
computed the temperature dependent correlation length for the attractive case and found
that the quasicondensation occurs also in this case. This is probably the most important
result of this Letter. We note that one does not expect gray solitons to appear in rapidly
cooled attractive Bose gas, so above mentioned connection to solitons certainly does not
hold in this case.
\begin{figure}
\includegraphics[width= 8.6cm, angle=0, clip=true]{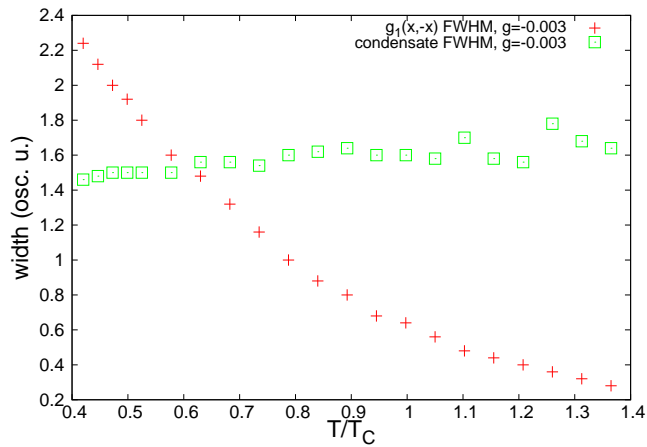}
\caption{(Color online) The width of the first order correlation function and the width of the condensate as functions of temperature.
}
\label{fig:quasi}
\end{figure}
The first order correlation function is defined as
\begin{equation}
 g_1 (x, -x) = \frac{\langle \Psi(-x)^* \Psi(x) \rangle}{\langle |\Psi(x)|^2 \rangle}
\label{eqn:g1}
\end{equation}
In Fig. \ref{fig:quasi} we plot its full-width at half maximum as a function of temperature and compare it
with the width of the condensate. The effect of quasicondensation is visible. It occurs at the
temperature $T\approx 0.6T_C$.

Summarizing: We have presented the first study of equilibrium
thermodynamics of the attractive 1D Bose gas trapped in a harmonic
potential. Our results are for the canonical statistical ensemble, thus
the temperature is a control parameter. The classical field approximation
is used. The appropriate probability distribution is obtained numerically
using a Monte Carlo technique. We found a reduced coherence length above
some characteristic temperature. This phenomenon was previously known only
for a repulsive gas.

\begin{acknowledgments}
\label{sec:acknowl}
This work was supported by Polish Government Funds for the years
2010-2012. 
Two of us (K.P. and K.Rz.) acknowledge 
financial support of the project  "Decoherence in long range interacting
quantum systems and devices" sponsored by the Baden-W\"urtenberg Stiftung".
\end{acknowledgments}

\end{document}